\documentclass[aps,prl,twocolumn,showpacs,groupedaddress,floatfix]{revtex4}

\usepackage{amsmath}
\usepackage{amsfonts}
\usepackage{amssymb}
\usepackage{graphicx}

\begin{document}

\title{Large anisotropy in the paramagnetic susceptibility of SrRuO$_{3}$
films}

\begin{abstract}
By using the extraordinary Hall effect in SrRuO$_{3}$ films we
performed sensitive measurements of the paramagnetic
susceptibility in this itinerant ferromagnet, from $T_{c}$ ($\sim
150$ K) to $300$ K. These measurements, combined with measurements
of magnetoresistance, reveal that the susceptibility, which is
almost isotropic at $300$ K, becomes highly anisotropic as the
temperature is lowered, diverging along a single crystallographic
direction in the vicinity of $T_{c}$. The results provide a
striking manifestation of the effect of exceptionally large
magnetocrystalline anisotropy in the paramagnetic state of a
\textit{4d} itinerant ferromagnet.
\end{abstract}

\author{Yevgeny Kats}
\author{Isaschar Genish}
\author{Lior Klein}
\affiliation{Department of Physics, Bar-Ilan University, Ramat-Gan 52900, Israel}
\author{James W. Reiner}
\altaffiliation{Present address: Department of Applied Physics, Yale University, New Haven,
Connecticut 06520-8284}
\author{M. R. Beasley}
\affiliation{T. H. Geballe Laboratory for Advanced Materials, Stanford University,
Stanford, California 94305}
\maketitle

The coupling of spin to electronic orbitals yields the ubiquitous phenomenon
of magnetocrystalline anisotropy (MCA) in ferromagnets \cite{MCA Kittel}.
While the manifestation of MCA below $T_{c}$ in the form of hard and easy
axes of magnetization is well studied, the fact that the strength of the MCA
decreases with temperature as a high power of the spontaneous magnetization
\cite{MCA vs T} could give the impression that MCA effects above $T_{c} $
are at most a weak perturbation. Here we show that the MCA has a significant
effect in the paramagnetic state of the \textit{4d}\ itinerant ferromagnet
SrRuO$_{3}$ over a wide range of temperatures -- not only is the
susceptibility diverging along a single axis at $T_{c}$, but the difference
between the susceptibilities along the different crystallographic axes is
noticeable (more than $30\%$) already at $t\equiv \left( T-T_{c}\right)
/T_{c}=0.5$.

A paramagnetic susceptibility diverging along only one
crystallographic direction has been reported previously for bulk
specimens of Cu(NH$_{4}$)Br$_{4}\cdot 2$H$_{2}$O \cite{Suzuki},
but the anisotropy there was found to be only $2\%$ of the
exchange integral $J$\ (compared to $20\%$ which we find in
SrRuO$_{3}$). Another report refers to two-dimensional cobalt
films, where an anisotropy of $5\%$ was measured \cite{Jensen}. In
both cases, the temperature range for which the susceptibility was
measured is by an order of magnitude smaller (in units of $T_{c}$)
than in our measurements of the three-dimensional SrRuO$_{3}$
films.

Measuring the paramagnetic susceptibility in films poses a
considerable technical challenge due to the combination of small
magnetic moment of the film with large background signal from the
substrate. We avoided these difficulties by using the
extraordinary Hall effect (EHE) whose signal depends on the film
internal magnetization and not on the total magnetic moment of the
sample. Therefore, the signal does not diminish with decreasing
thickness, neither is it affected by the substrate magnetization.

We study high-quality epitaxial films of the itinerant ferromagnet
SrRuO$_{3} $ ($T_{c}\sim150$ K)\ with thicknesses in the range of
$6$-$150$ nm patterned by photolithography for measurements of
resistivity and Hall effect. The unit cell is orthorhombic
($a=5.53\mathrm{\ }$\textrm{\AA }, $b=5.57\mathrm{\ }$\textrm{\AA
}, $c=7.85\mathrm{\ }$\textrm{\AA }), and the single easy axis of
magnetization is roughly in the $b$ direction \cite {TEM,Mag SRO}.
The films are\ grown by reactive electron beam coevaporation
\cite{films growth} on miscut ($\sim2^{\circ}$) SrTiO$_{3}$\
substrates. This technique produces single-phase films with the
[110] direction perpendicular to the surface (as was shown by
transmission electron microscopy study of films grown in the same
apparatus \cite{TEM}), so that the $b$ direction is at
$45^{\circ}$ out of the plane of the film.

The transverse electric field $\mathbf{E}_{H}$\ in magnetic
conductors originates from both the \textit{ordinary} (or
\textit{regular}) Hall effect (OHE), which depends on the magnetic
induction $\mathbf{B}$, and the \textit{extraordinary} (or
\textit{anomalous}) Hall effect (EHE), which depends on the
magnetization $\mathbf{M}$:
\begin{equation*}
\mathbf{E}_{H}=-R_{0}\mathbf{J\times B}-R_{s}\mathbf{J\times }\mu _{0}%
\mathbf{M}\text{,}
\end{equation*}%
where $\mathbf{J}$ is the current density,\ $R_{0}$ is the ordinary Hall
coefficient related to the carrier density $n$, and $R_{s}$ is the
extraordinary Hall coefficient whose temperature dependence in the
ferromagnetic phase of SrRuO$_{3}$\ has been reported in Refs. \cite{EHE
Izumi, EHE Klein}.

\begin{figure}[ptb]
\includegraphics[scale=0.5, trim=125 420 200 -180]{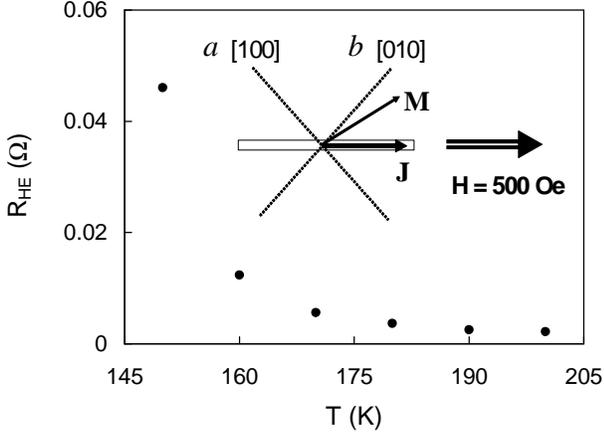}
\caption{Hall effect with $H=500$ Oe applied parallel to the
current (along the [$1\overline {1}0$] direction) as a function of
temperature above $T_{c}$ ($\simeq147$ K).} \label{in-plane EHE}
\end{figure}

In measurements above $T_{c}$ we found that a significant
Hall effect develops even when the magnetic field is applied
parallel to the current which flows along the [$1\overline{1}0$]
direction. The temperature dependence of this Hall effect
resembles the behavior of the induced magnetization (see Fig.
\ref{in-plane EHE}). These results indicate that the in-plane
field generates a significant out-of-plane component of
$\mathbf{M}$, resulting in a measurable EHE. This implies that the
paramagnetic susceptibility of SrRuO$_{3}$ films is described by
an anisotropic tensor.

\begin{figure}[tbp]
\includegraphics[scale=0.44, trim=150 320 120 -190]{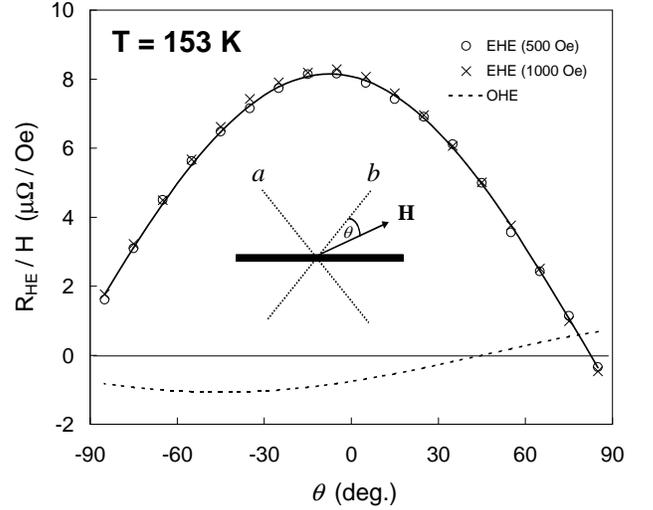}
\caption{EHE at $T=153$ K divided by the applied field $H$
(circles: $500$ Oe, crosses: $1000$ Oe) as a function the angle
$\theta$\ between $\mathbf{H}$ and the $b$ direction (see
illustration). The dashed curve is the OHE. The solid curve is a
fit obtained by assuming certain constant values for the
susceptibility along the $a$ and $b$ directions:
$R_{s}\protect\chi _{a}=0.48\times 10^{-9}$ $\Omega \,$m/T,
$R_{s}\protect\chi _{b}=3.60\times 10^{-9}$ $\Omega \,$m/T.}
\label{low-field EHE}
\end{figure}

For quantitative characterization of the susceptibility
anisotropy, we measured the Hall effect as a function of field
direction at various temperatures. For each temperature above
$T_{c}$, a small-field limit exists, where the magnetization
depends linearly on the field and can be fully described in terms
of constant susceptibilities $\chi _{a}$, $\chi _{b}$, and $\chi
_{c}$ along the $a$, $b$, and $c$ crystallographic directions,
respectively ($\mu _{0}M_{a}=\chi _{a}H_{a}$, etc). An example of
measurements in this limit is shown in Fig. \ref{low-field EHE},
where the EHE resistance ($R_{EHE}=\mu_{0}R_{s}M_{\bot }/t$, where
$t$ is the thickness of the sample) is shown for two different
fields at $T=153$ K as a function of the angle $\theta $ (see
inset). The solid curve is a fit obtained by assuming certain
values of $\chi _{a}$ and $\chi _{b}$, based on the equation:
\begin{equation*}
R_{EHE}\left( H,\theta \right) =\frac{R_{s}H}{\sqrt{2}t}\left(
\chi _{b}\cos \theta -\chi _{a}\sin \theta \right) \text{.}
\end{equation*}
Figure \ref{low-field EHE} also demonstrates the relatively small
magnitude and different angular dependence of the OHE, which was
subtracted from the measured signal \cite{OHE}.

\begin{figure}[tbp]
\includegraphics[scale=0.46, trim=125 345 210 -155]{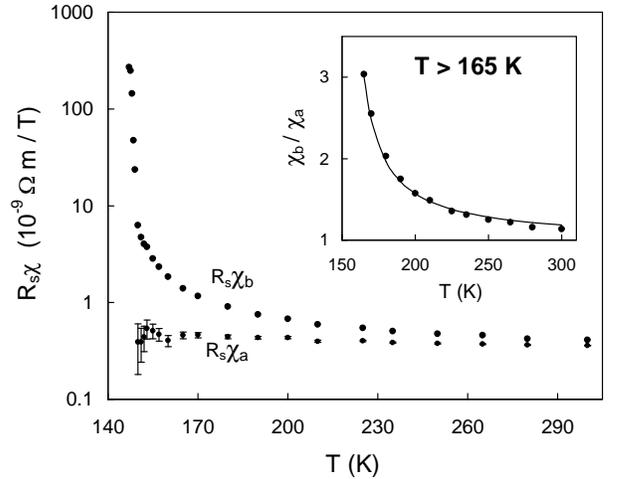}
\caption{Susceptibility along the crystallographic directions
[100] ($\chi _{a}$) and [010] ($\chi _{b}$) as a function of
temperature, on a semilog plot. The values are multiplied by
$R_{s}$, whose temperature dependence is expected to be smooth.
The error bars for $R_{s}\chi _{a}$\ reflect an uncertainty of up
to $2^{\circ}$ in $\theta$. The inset shows the ratio
$\chi_{b}/\chi _{a}$ for $165$ K $<T<$ $300$ K. The solid curve is
a fit to $\left( T-T_{c,a}^{MF}\right) /\left(
T-T_{c,b}^{MF}\right) $ with $T_{c,a}^{MF}=123$ K,
$T_{c,b}^{MF}=151$ K.} \label{RsX vs T}
\end{figure}

The main result of this Letter is presented in Fig. \ref{RsX vs
T}, which shows the temperature dependence of the susceptibilities
$\chi _{a}$ and $\chi _{b} $ (multiplied by $R_{s}$). We see that
the susceptibility is very anisotropic throughout most of the
investigated temperature range. Particularly, $\chi _{b}$ exhibits
striking divergence at $T_{c}$, while $\chi _{a}$ changes
moderately. The actual divergence of $\chi _{b}$\ is even stronger
than shown in Fig. \ref{RsX vs T} since $\chi _{b}$ was not
corrected for the demagnetizing field (because of uncertainty in
the value of $R_{s}$); consequently, the apparent susceptibility
in our measurement configuration is only $\chi _{b}/\left( 1+\chi
_{b}/2\right) $.

\begin{figure}[tbp]
\includegraphics[scale=0.5, trim=170 375 160 -145]{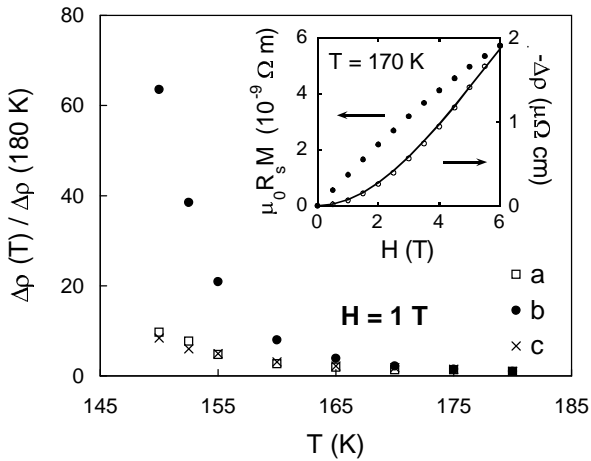}
\caption{Magnetoresistance as a function of temperature with a
field of $1$ T applied along the different crystallographic
directions (the values are
normalized to the values at $180$ K). The inset shows the magnetization ($%
\protect\mu _{0}R_{s}M$) and magnetoresistance ($\Delta \protect\rho $) as a
function of a field (applied along the $b$ direction, at $T=170$ K). The
solid curve is a fit to $\Delta \protect\rho \propto -M^{2}$.}
\label{M MR}
\end{figure}

Since the $c$ direction is in the plane of the film, the EHE
measurement could not be used to determine $\chi _{c}$ (the
insensitivity of EHE to a field component in the $c$ direction was
experimentally confirmed). Therefore, measurements of
\textit{magnetoresistance} (MR)\ $\Delta \rho \equiv \rho \left(
H\right) -\rho \left( 0\right) $\ were employed \cite{MR in SRO}.
Based on previous \cite{MR SRO}\ and current results (see inset to
Fig. \ref{M MR}), $\Delta \rho \propto -M^{2}$ (for a constant
direction of magnetization). Thus we can infer the susceptibility
behavior along $a$, $b$, and $c$ directions by comparing the MR
obtained with fields applied along these directions. The results,
shown in Fig. \ref{M MR}, clearly indicate that the induced
magnetization along the $b$ direction grows as $T_{c}$ is
approached much more rapidly than along the $a$ or $c$ directions.
The divergence here is less pronounced than in Fig. \ref{RsX vs T}
since for fields applied here the magnetization along $b$ is
sub-linear (but using lower fields would not allow obtaining
accurate MR data for the $a$ and $c$ directions). The temperature
dependence of the MR with $\mathbf{H} \parallel c$ is very similar
to the MR with $\mathbf{H} \parallel a$, suggesting that the
behavior of $\chi _{c}$ is similar to the behavior of $\chi _{a}$.

The data in Figs. \ref{RsX vs T} and \ref{M MR} imply that only
the susceptibility along the $b$ direction (which is also the easy
axis of the spontaneous magnetization \cite{TEM,Mag SRO})\
diverges at the phase transition; moreover, the large anisotropy
of the susceptibility is noticeable ($>30\%$) already at $t\equiv
\left( T-T_{c}\right) /T_{c}=0.5$.

We note that this result is consistent with a previous report on
Ising-like critical behavior in SrRuO$_{3}$ films \cite{MR SRO}
and with the possibility to fit specific heat data of SrRuO$_{3}$
crystals with Ising exponent \cite{MF SRO}; an alternative
analysis of the critical behavior given in Ref. \cite{MF SRO} does
not seem to be consistent with the results reported here.

Anisotropy in the behavior of the susceptibility arising from MCA
may be described microscopically by Heisenberg model with
\textit{anisotropic exchange}:
\begin{equation}
\mathcal{H}=-\sum\limits_{<ij>,\alpha }J_{\alpha }S_{i\alpha
}S_{j\alpha }
\label{anisotropic exchange}
\end{equation}
or with \textit{single-site anisotropy}:
\begin{equation}
\mathcal{H}=-J\sum\limits_{<ij>}\mathbf{S}_{i}\mathbf{\cdot
S}_{j}-\sum\limits_{i,\alpha }D_{\alpha }S_{i\alpha }^{2} \text{,}
\label{single-site anisotropy}
\end{equation}
where $\alpha =a,b,c$ denotes spin components along the
crystalline directions, and $i$, $j$ denote lattice sites. Band
calculations \cite{BC SRO} and spin polarization measurements
\cite{SP SRO} show that SrRuO$_{3}$ is an itinerant ferromagnet.
However, various theoretical models (e.g., the local-band theory
\cite{local-band theory}) indicate that magnetic moments in
itinerant ferromagnets can behave as if localized even above
$T_{c}$, thus vindicating the description of their magnetic
interactions by Heisenberg Hamiltonian. The experimental
observation that the exchange splitting in SrRuO$_{3}$ does not
change much as $T_{c}$ is approached \cite{exchange splitting}
supports the relevance of such a treatment to SrRuO$_{3}$.

Since $T_{c}\propto J$, anisotropic exchange results in a
different effective $T_{c}$ for each spin component. Consequently,
only the susceptibility along the direction with the largest $J$
diverges at the actual $T_{c}$ \cite{Pfeuty Toulouse book}.
Single-site anisotropy yields the same critical behavior
\cite{Pfeuty Toulouse book}, with an effective $\Delta J=\Delta
D/z$, where $z$ is the number of nearest neighbors.

The anisotropy $\Delta J$ (or $\Delta D$) can be estimated from
the susceptibilities in the mean-field region, which are expected
to follow the Curie-Weiss $1/\left( T-T_{c}^{MF}\right) $ law
(with different mean-field transition temperature $T_{c}^{MF}$ for
each direction), if Pauli paramagnetism and diamagnetism can be
neglected \cite{Pauli PM}. We cannot fit $R_{s}\chi _{a}$ or
$R_{s}\chi _{b}$ as a function of temperature because the
temperature dependence of $R_{s}$ is unknown. However, the ratio
$\chi _{b}/\chi _{a}$ does not depend on $R_{s}$, and fitting the
data to the expression $\left( T-T_{c,a}^{MF}\right) /\left(
T-T_{c,b}^{MF}\right) $ converges to the values
$T_{c,a}^{MF}=123\pm 2$ K, $T_{c,b}^{MF}=151\pm 1$ K for $165$ K
$<T<$ $300$ K (see inset in Fig. \ref{RsX vs T}). From these
values we find an anisotropy of $J_{b}-J_{a}\simeq 0.2J_{avg}$ or
$D_{b}-D_{a}\simeq 0.2Jz$. This result is almost
thickness-independent from $6$ to $150$ nm (the data presented
above are from a $30$-nm film).

To examine whether the paramagnetic anisotropy is consistent with
the ferromagnetic anisotropy we performed magnetization
measurements below $T_{c}$. However, noting that the easy axis of
the magnetization, which is along the $b$ direction close to
$T_{c}$, changes its orientation as the temperature is lowered (by
a maximum of $15^{\circ}$ at zero temperature)
\cite{reorientation} and considering that additional factors
become significant when the magnetization is large, the most we
can expect is an order-of-magnitude agreement between the
paramagnetic and the ferromagnetic anisotropy.

Using a SQUID magnetometer which measures the whole magnetization
vector we estimated the ferromagnetic anisotropy by measuring the
rotation of $\mathbf{M}$ resulting from applying a magnetic field
in the $[1\overline{1}0]$ direction. We find, between $10$ and
$90$ K, a smooth rotation of $\mathbf{M}$ in the $(001)$ plane by
up to $\sim20^{\circ}$ at $H=5$ T (see Fig. \ref{Ferromagnetic
anisotropy}) almost without change of magnitude \cite{substrate
M}. This behavior can be described by an anisotropy energy,
$E_{anis} = K \sin ^2\theta$, with a weakly-temperature-dependent
anisotropy constant $K$ whose low-temperature value is
$(1.2\pm0.1)\times 10^{7}$ erg/cm$^3$.

\begin{figure}[tbp]
\includegraphics[scale=0.5, trim=140 395 150 -110]{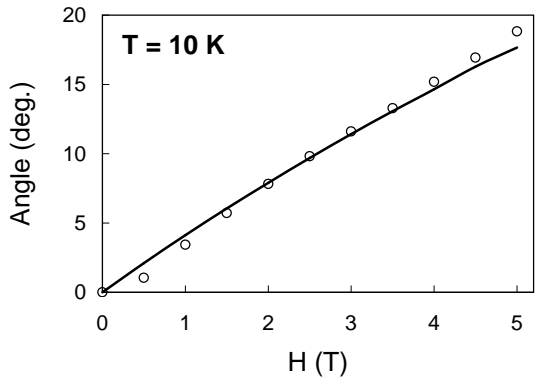}
\caption{Deviation of the magnetization from the easy axis as a
function of the field which is applied at $60^{\circ }$ relative
to the easy axis. The solid curve presents the expected behavior
with an anisotropy constant of $1.2 \times 10^{7}$ erg/cm$^{3}$.}
\label{Ferromagnetic anisotropy}
\end{figure}

The large anisotropy constant (compared to $5\times 10^{5}$
erg/cm$^{3}$ in Fe \cite{anis Fe}, $8\times 10^{5}$ erg/cm$^{3}$
in Ni \cite{anis Ni}, and $4\times 10^{6}$ erg/cm$^{3}$ in hcp Co
\cite{anis Co}) is probably a result of the reduced symmetry
(orthorhombic) \cite{symmetry} and the large spin-orbit coupling
\cite{SOC} through which the spin direction is affected by the
crystal structure.

To relate the ferromagnetic anisotropy with Eqs. (\ref{anisotropic
exchange}) and (\ref{single-site anisotropy}), we note that at low
temperature the exchange aligns the spins along a single
direction, thus the energy cost of magnetization rotation from the
$b$ direction toward the $a$ direction by an angle $\theta $ in
the case of anisotropic exchange is $\Delta
E=zNS^{2}(J_{b}-J_{a})\sin ^{2}\theta $, where $N$ is the number
of spins per unit volume. A similar result is obtained in the case
of single-site anisotropy. Considering that the zero-temperature
magnetization is $1.4\mu_{B}$ per Ru ion, and calculating $J$
according to the relation $J=3k_{B}T_{c}^{MF}/2zS\left(
S+1\right)$ we obtain $J_{b}-J_{a} \simeq 0.1J_{avg}$. This result
is in reasonable agreement with $J_{b}-J_{a} \simeq 0.2J_{avg}$
extracted from the anisotropic susceptibility.

Finally we would like to note that while it is common to use the
EHE as an indicator of magnetization, it is seldom used for a
quantitative analysis of magnetic behavior. In this work we
presented a striking example of the latter, by performing
sensitive measurements of the zero-field-limit magnetic
susceptibility in thin films of SrRuO$_{3}$, which allowed us to
gain microscopic insight into the effect of MCA in the
paramagnetic state of a \textit{4d}\ itinerant ferromagnet.

We would like to thank V. Kosoy for an illuminating discussion,
and A. Aharony, J. S. Dodge, and M. Gitterman for useful comments
on the manuscript. We acknowledge support by the Israel Science
Foundation founded by the Israel Academy of Sciences and
Humanities.


\begin{thebibliography}{99}
\bibitem{OHE} OHE was subtracted based on the value of $R_{0}$ determined
for each sample by measuring the Hall effect at a temperature where $R_{s}$
vanishes, close to $130$ K (we could determine that temperature up to the
desired accuracy by the methodology of Ref. \cite{EHE Klein}). While small
changes in $R_{0}$\ as a function of temperature cannot be excluded, our
main results would be affected very little by such changes.
\bibitem{MR in SRO} The MR in SrRuO$_{3}$ is negative and related to
suppression of spin-dependent scattering. Only at very low
temperatures positive Lorentz magnetoresistance appears.
\bibitem{Pauli PM} The Pauli paramagnetism and diamagnetism in SrRuO$_{3}$\
are approximately $10\%$ of the susceptibility at $300$ K
\cite{Cao}. It is not clear whether their contribution to the EHE
should go with the same $R_{s}$. If same $R_{s}$ is assumed,
taking them into account does not have a significant effect on the
result or the quality of the fit.
\bibitem{reorientation} This feature has been reported previously
\cite{Mag SRO}, and it does not depend on film thickness. It is
not due to the shape anisotropy, which is small ($M_{s}^{2}
\approx 5 \times 10^{4}$ erg/cm$^{3}$ $\ll K$).
\bibitem{substrate M} The contribution of the substrate was subtracted based
on measurements of a bare substrate. Its signal is mainly
diamagnetic, being for $H=5$ T about $20\%$ of the ferromagnetic
signal of the $150$-nm film which was measured.
\bibitem{symmetry} A remarkable example of the dependence of MCA
on the crystal symmetry is the order-of-magnitude difference in
the anisotropy constants bewteen hcp and fcc Co \cite{anis Co}.
\end{thebibliography}
\end{document}